\documentclass{elsarticle}
\usepackage[latin9]{inputenc}
\usepackage{textcomp}
\usepackage{amsmath}
\usepackage{amssymb}
\usepackage{graphicx}
\usepackage{esint}
\usepackage[english]{babel}

\makeatother

\usepackage{array}
\usepackage{textcomp}
\usepackage{multirow}
\makeatletter
\providecommand{\tabularnewline}{\\}

\usepackage{babel}
\usepackage{babel}
\usepackage{babel}
\usepackage{babel}
\usepackage{babel}
\usepackage{babel}
\usepackage{babel}\makeatother
\usepackage{babel}

\begin{document}

\title{Correlations in energy in cosmic ray air showers radio-detected by
CODALEMA}

\author[add1]{Ahmed Rebai}
\author[add1]{Pascal Lautridou}
\author[add2]{Alain Lecacheux}
\author[add1]{Olivier Ravel}

\address[add1]{SUBATECH, IN2P3-CNRS/Université de Nantes/Ecole des Mines de Nantes,
Nantes, France.}

\address[add2]{LESIA, USN de Nançay, Observatoire de Paris-Meudon/INSU-CNRS, Meudon,
France.}

\begin{abstract}
A study of the response in energy of the radio-detection method of
air showers initiated by ultra-high-energy cosmic rays is presented.
Data analysis of the CODALEMA experiment shows that a strong correlation
can be demonstrated between the primary energy of the cosmic ray and
the electric field amplitude estimated at the heart of the radio signal.
Its sensitivity to the characteristics of shower suggests that energy
resolution of less than 20\% can be achieved. It suggests also that,
not only the Lorentz force, but also another contribution proportional
to all charged particles generated in the development of the shower,
could play a significant role in the amplitude of the electric field
peak measured by the antennas (as coherence or the charge excess). 
\end{abstract}
\maketitle

\section{Introduction}

So far, the study of Ultra-High-Energy Cosmic rays (UHECR) has relied
primarily on the use of particle detectors and fluorescence detectors.
Today, further improvements are envisaged by supplementing the conventional
measures of Extensive Air Showers (EAS). One of them exploits the
detection of the transient radio signal induced by the movement of
secondary charged particles in the atmosphere. This method is not
novel; it emerged in the early 60s, was abandoned in the mid-70s \cite{key-1}
and then re-launched in early 2000. Several experiments (such as Lopes
\cite{key-2}, CODALEMA \cite{key-3}, and many new developments\cite{key-4}),
supported by renewed theoretical foundations \cite{key-5,key-6,key-7,key-8,key-9,key-10,key-11},
are now re-investigating the method. Up to now, the techniques of
detection and identification of the shower, of reconstruction of arrival
directions, of topology of the electric field , and of emission mechanisms
have been most often discussed in the literature \cite{key-12,key-13,key-14,key-15},
while the problem of estimating the energy through the radio method
was, up to now, only briefly been touched on in the publications \cite{key-16,key-17}.

In this work, the emphasis will be on how and with what, a radio-estimator
of the energy of the showers can be derived. We shows that, subject
to adjustments related to the mechanism of radio emission, a robust
correlation can be demonstrated between the energy of a cosmic ray
derived from the analysis of the distribution of the particles at
ground level and the electric fields measured by antennas, and that
it provides an energy estimator by the radio method. Section 2 recalls
the essential features of the experimental devices, paying attention
to the analysis of the particle detectors and to the derivation of
the energy of the showers using the CIC method \cite{key-18}. The
radio-detection performances and the essential features of the showers
used for the energy comparison are central to Section 3. Section 4
discusses how, from the analysis of the amplitudes of the signals
measured by the antenna array, it is possible to extract an observable
-- the amplitude of the electric field at its center -- which depends
on the EAS primary energy. The conclusions appear in the last section.

\section{Principle of the experiment}

\subsection{Detector set-up}

The CODALEMA experiment is installed at the radio observatory of Nançay,
France. Detailed descriptions of the set-up and of the analysis methods
of CODALEMA have already been given in earlier publications \cite{key-19,key-20,key-21}. 

The radio set-up consists of an array of $24$ short active dipoles
distributed on a cross of $600\times500\; m$. It provides an independent
measurement of radio-electric field waveforms in an E-W polarization
on each antenna (except for $3$ that sample in a N-S polarization).
The sensor is based on an active dipole using a wide and short radiator
\cite{key-22}. It operates in a frequency bandwidth from $1\: MHz$
to $200\: MHz$ and is sensitive to the galactic noise background
\cite{key-23}. (A second radio-system, the Nançay decameter array,
is dedicated to the study of the electric field with a high spatial
resolution \cite{key-24} using $144$ tepee antennas, but its results
are not discussed in the current study).

The particle detector array is made up of $17$ plastic scintillator
stations, which are derived from those used by the EAS-TOP experiment
of Gran Sasso \cite{key-25,key-26}. These stations are spread over
a square grid of $340\times340\; m$ with a spacing of $80\: m$.
The station houses $2$ photomultipliers (PMT), one working at high
gain, the other at low gain, so as to provide an overall dynamics
of $0.3$ to $3000$ Vertical Equivalent Muons (VEM); this unit corresponds
to the mean charge deposited by a single muon crossing the scintillator
plate vertically. The detection reaches the maximum efficiency for
an energy of $10^{16}\: eV$.

The trigger for all the system is provided by the coincident detection
of the five central particle detectors leading to an energy detection
threshold located around $10^{15.7}\: eV$, at the ``knee'' energy.
For each trigger signal, all waveforms of all detectors are digitized
(sampling rate of $1\: GS/s$, duration of $2.56\:\mu s$ and recorded.
All the sensors are connected to the central acquisition room through
coaxial cables.

\subsection{Event selection}

The off-line analysis is conducted independently for the two detector
arrays. 

Concerning the radio device, the radio transient is searched for in
the antenna waveform through a $24-82\: MHz$ frequency bandwidth
where the electric field peak value and the arrival time are extracted.
If more than 3 radio transients are detected in coincidence, the arrival
direction of the electric field (zenith angle $\theta$, azimuth angle
$\phi$) is reconstructed by considering a planar wavefront. Assuming
a decrease in the electric field profile of the form $\varepsilon_{0}\exp(-d/d_{0})$,
four new observables are introduced: the electric field magnitude
at the radio shower center $\varepsilon_{0}$ , the slope parameter
of the electric field $d_{0}$ , and the impact location of the radio
shower center on the ground ($x_{0},y_{0}$). During this processing,
the standard deviation of the noise on each antenna is used as the
measurement error on the electric field. The timing error is estimated
to 10 ns, leading to an angular accuracy of reconstructed arrival
directions in the range of the degree \cite{key-27}. 

The goal of the particle detectors is to provide the arrival direction,
the size, the core location of the EAS and an estimation of its energy,
which will serve as a reference for the current study. The arrival
direction of the particle front (assumed planar) is estimated from
the relative arrival time in each particle detector. An analytical
NKG lateral distribution \cite{key-28} is then adjusted on the measured
particle densities in the shower frame. If the shower core position
falls inside the particle detector array, the event is referred to
as internal and the fitted NKG distribution is used to estimate the
primary energy using the constant intensity cut (CIC) method \cite{key-29}.

The features of these events, \textquotedblleft{}radio\textquotedblright{}
and \textquotedblleft{}particles\textquotedblright{}, are then used
to identify the coincidences. In this work, the data collected by
CODALEMA during $659$ days from November 27, 2006 to March 6, 2009
were analyzed. During this period, $143795$ triggers were registered.
The radio-detected showers were selected according to several successive
criteria \cite{key-27}. Only events ``radio'' with at least $4$
flashed antennas and for which the coincidence, below $100\: ns$
in time and $20{^{\circ}}$ in arrival direction, between the wavefront
``particle'' and ``radio'' were retained. These constraints reduced
the available statistics to $1386$ coincidences. To benefit from
an energy estimation as reliable as possible, only ``particles''
events with footprints located inside the particle detector array,
with zenith angles lower than $60{^{\circ}}$ and energies above $10^{16.7}\: eV$,
were kept (events referred to as ``internal''). At the end of this
selection, only $376$ coincidences were sufficiently reliable to
be used in further analysis. These criteria, translated in energy
spectra and in the efficiency curve of radio-detection, are presented
in Fig. 1.

\begin{figure} 
\centering
\includegraphics[width=7cm,height=7cm]{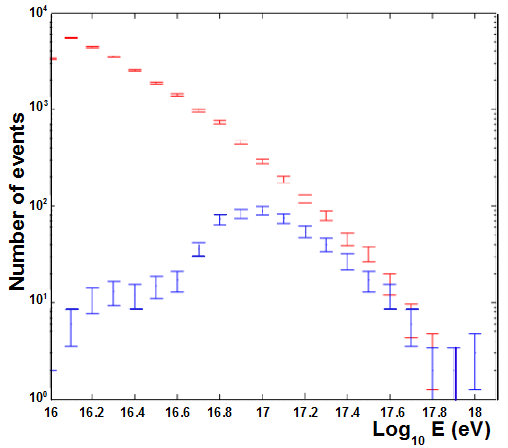}\includegraphics[width=7cm,height=7cm]{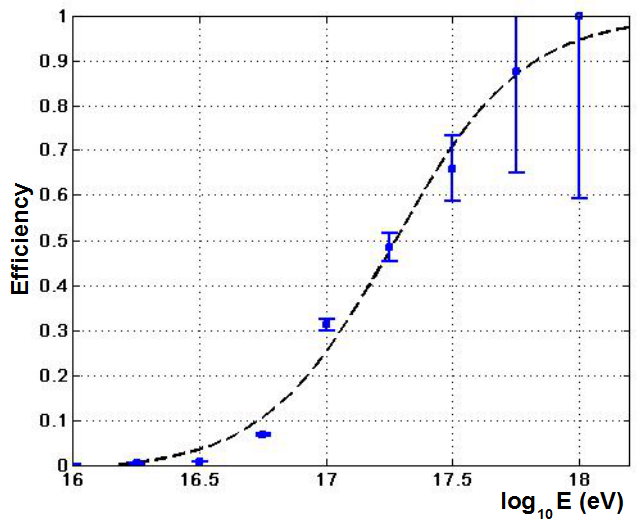}

\caption{Left: Full energy distribution seen by the particle detector array
(red markers) and the subset in coincidence with the radio antenna
array (blue markers). Right: ``Radio'' efficiency (in the east-west
polarization state) versus the energy given by the particle detector
array.}

\end{figure}

\subsection{Energy estimation with the particle detectors}

The estimation of the shower energy for the particle detectors, relies
fundamentally on the information on the deposited charge, and on the
time, from the digitized waveform recorded at the trigger occurrence.
For this, the main features of the PMT signal (timing $t_{0}$ and
amplitude $V_{max}$, are first obtained by fitting it to the following
shape:

\[
V(t)=V_{0}+V_{max}\left(\frac{(t-t_{0})e}{n\tau}\right)^{n}e^{-\frac{t-t_{0}}{\tau}}
\]

where $V_{0}$ is the background baseline, which could be taken as
constant in the fit interval. The corresponding charge Q is evaluated
from the resulting fitted function $V(t)$. For each station, this
charge is converted to a VEM number, by calibration from specific
counting of the single muons. The VEM number, corresponding to the
average value of the measured spectrum corrected by the mean crossing
angle $36.9{^{\circ}}$, serves as calibration factor to convert the
charge Q into the equivalent particle density on each detector.

In parallel, the zenith angle $\theta$ and azimuth angle $\phi$
of the shower are obtained by a planar adjustment from the relative
time of flight between the different stations. Both angular resolutions
(on $\theta$ and $\phi$) are of the order of $1{^{\circ}}$ for
events of energy higher than a few $10^{16}\: eV$, noting that both
associated errors are strongly correlated.

Event by event, station positions are projected into the plane perpendicular
to the arrival direction and containing the central detector. A MINUIT
minimization algorithm, assuming an NKG theoretical shape, performs
the reconstruction of the lateral distribution of the particle densities.
The core position is projected back into the array plane. Internal
events are identified by asking for a larger particle density in internal
detectors than in surrounding ones. The total internal surface area
of the apparatus is then used for energy and flux estimations.

The estimation of the incident particle energy is performed in two
different steps: 

The Constant Intensity Cut method (CIC) is used to infer the attenuation
length $\Lambda_{att}$ which expresses the decrease in the shower
size $N_{e}(\theta)$ with increasing zenith angle. For each event
with $N_{e}(\theta)$, an equivalent vertical shower size $N_{e}(0)$
is extracted from the measured $N_{e}$ and $\theta$. 

Vertical air shower simulations with the AIRES Monte-Carlo code \cite{key-6},
at different energies and for incident protons, provide the correspondence
between the primary energy $E_{proton}$ and the mean value of the
simulated shower sizes $<N_{e}(0)>$ at each energy $E_{proton}$. 

The usual integral intensities $J(>N_{ec},\theta)$ (in $m^{-2}.s^{-1}.sr^{-1}$)
\cite{key-30} are calculated for a shower size greater than a given
value $N_{ec}$. The corresponding attenuation curves $N_{e}(\theta)$
are derived, as well as the corresponding attenuation lengths $\Lambda_{att}$.
The mean experimental value of $\Lambda_{att}$ ($188\, g/cm^{2}$)
is very close to that extracted from AIRES simulations ($190\, g/cm^{2}$).
The incident energy $E\:(eV)$ is deduced from AIRES simulations,
by identifying $N_{e}(0)$ to the mean value $N_{e}(0)$ obtained
from one hundred shower simulations at a given energy and a given
kind of primary ray. Six proton energies in the range $10^{16}-10^{18}\: eV$
have been considered, leading to the energy estimation:

$E_{p}(eV)=2.138\:10^{10}\,<N_{e}(0)>^{0.9}$

The main uncertainty about the energy estimation is due to the shower
to shower fluctuations at a fixed energy and is of the order of $30\,\%$
at $10^{17}\, eV.$ The primary flux, extracted from the data, has
been compared to published results in this energy range (Fig. 2).
The results are very close to the proton curve, although our setup
is probably not sophisticated enough to draw strong conclusions from
this result.

\begin{figure}
\centering
\includegraphics[width=14cm,height=7cm]{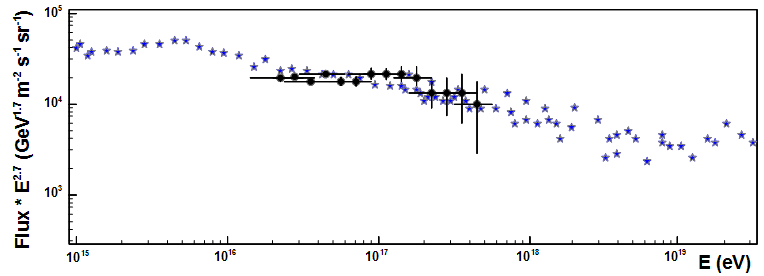}

\caption{Measured flux $E^{2.7}\times\Phi(E)\;[GeV^{1.7}m^{-2}s^{-1}sr^{-1}]$
as a function of $E\:(eV)$. The black circles are our calculations
for proton primaries. The blue stars are from cosmic ray data compilations.}
\end{figure}

\section{Analysis of the radio data}

\subsection{Data processing of the signal radio}

As this study relies on the analysis of the amplitude of the transient
electric field received by the antenna, during the data processing
several corrections are applied to estimate the actual magnitude of
the incident electric field. Taking into account these effects, a
conversion of the measured voltage amplitude (in $V$) to an electric
field (in $V/m$) detected by the antenna is achieved.

A numerical filtering of the waveform ($24-82\: MHz$ band pass filter)
is undertaken in order to eliminate the contributions of AM and FM
radio transmitters.

A compensation function versus the frequency is introduced to take
into account the measured signal attenuation along the length of the
cable (up to $380\: m$) carrying the signal from the antenna to the
digitizer and insertion losses due to the connectors.

As the directivity of the radio reception of the dipole located near
the ground is not isotropic, a correction of the peak amplitude of
the transient signal, based on the arrival direction of the EAS, is
applied. This is achieved using the output results of the EZNEC simulation
software \cite{key-31} for the radiation pattern of the antenna.

In contrast, the effect of the group velocity of the signal during
its propagation through the chain of detection, was ignored. Due to
the absence of analog filters in our electronic chain, such a correction
is not considered critical in the $24-82\: MHz$ band used in this
study.

Given the speed of sampling of the waveform digitizer ($1\, GS/s$,
analog bandwidth $0-300\: MHz$) and the frequency band used ($24-82\: MHz$),
the AD conversion introduces a non-significant deformation of the
peak amplitude and no correction was applied.

The signal deformations due to the environment near the sensors (trees,
buildings) were also searched for, but analysis did not reveal an
important role for this parameter \cite{key-32}.

\subsection{Features of selected ``radio'' events}

In practice, the only available knowledge to construct the radio observables
is the measured arrival time $t_{i}$ and the strength $\varepsilon_{i}$
of the radio transients, which are picked up at different locations
of the radio-detection area (at antenna positions $x_{i}$, $y_{i}$).
These observables are thus independent of those deduced from the particle
detector array.

In agreement with historical observations \cite{key-1} and recent
theoretical developments \cite{key-5,key-6,key-7,key-8,key-9,key-10,key-11},
we used, event by event, an exponential Radio Lateral Distribution
Function (RLDF). It describes the radio signal amplitude as a function
of the distance d to the shower axis in the frame of the shower, as
an exponential law of the form (Allan's parametrization):

\begin{equation}
\varepsilon_{d}=\varepsilon_{0}\exp\left(-\frac{d}{d_{0}}\right)
\end{equation}

With our convention of angles (azimuthal angles $\phi=0\text{\textdegree}$
to the north and $\phi=90{^{\circ}}$ to the west, zenith angles $\theta$)
and because the detectors are located at the same altitude, this relation
can be re-written as:

\begin{equation}
\varepsilon_{d,i}=\varepsilon_{0}\exp\left(-\frac{[(x_{i}-x_{0})^{2}+(y_{i}-y_{0})^{2}-((x_{i}-x_{0})\cos\theta\sin\phi+(y_{i}-y_{0})\sin\theta\sin\phi)^{2}]^{1/2}}{d_{0}}\right)
\end{equation}

This simplest approach introduces the 4 new ground observables for
each shower: $(\varepsilon_{0},\, d_{0})$ and $(x_{0},\, y_{0})$,
respectively the electric field magnitude at the core, the slope parameter
of the electric field (the speed at which the transient electrical
signal decreases with the distance to the axis), and the impact location
of the shower on the ground.

The functional is estimated event by event, providing the coordinates
$(x_{i},\, y_{i})$, the signal amplitude $\varepsilon_{i}$ of the
tagged antennas, and the zenith and azimuth angles $(\theta,\,\phi)$
of the radio wavefront reconstructed by a planar fit (which itself
uses the arrival time measurements ti). The fit of the RLDF is computed
only for showers having at least $5$ tagged antennas, as a consequence
of the four free parameters. Figure 3 presents an RLDF reconstructed
from a high antenna multiplicity event.

\begin{figure}
\centering
\includegraphics[width=7cm,height=7cm]{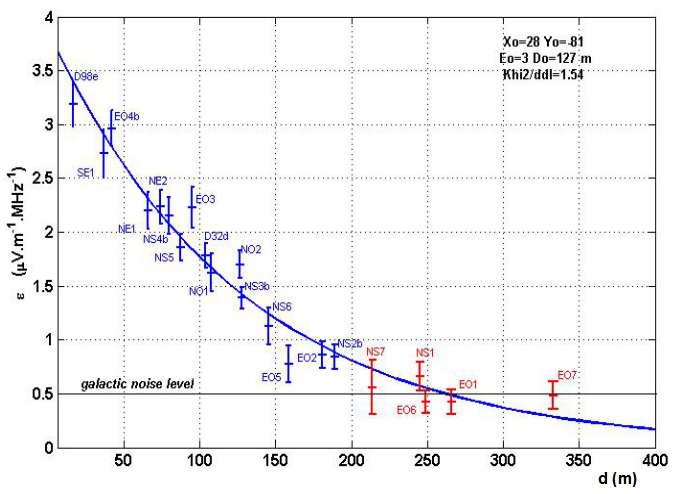}

\caption{Result of the fit of an RLDF of the form $\varepsilon_{0}\exp(-d/d_{0})$
for a radio event observed by CODALEMA in the referential of the shower
''radio''.}
\end{figure}

One of the most delicate aspects of the adjustment is to assign an
error to the observable $\varepsilon_{0}$ .This is achieved using
the selected ``radio'' event as an input of a Monte Carlo calculation.
For each event, a thousand RLDF fits are repeated, with signal amplitudes
$\varepsilon'_{i}$ randomly taken from a Gaussian probability density
function centered on the measured values $\varepsilon_{i}$ , within
$\pm\,3\sigma$ for each tagged antenna. This error is defined as
the standard deviation of the measured radio noise on each antenna.
During the convergence of the $\chi^{2}$ process, the observables$(\varepsilon_{0},\, d_{0})$
and core position $(x_{0},\, y_{0})$ are left free while the arrival
direction $(\theta,\,\phi)$ and the antenna positions stay fixed.
This procedure enables the relevant part of the phase space to be
explored and thus the error $\varepsilon_{0}$ to be estimated on
an event by event basis. The results are combined in Fig. 4. By choosing
a functional error similar to that used for the particles (not per
event), on average a $22\:\%$ error affects the estimation of $\varepsilon_{0}$
\cite{key-33}.

\begin{figure}
\centering
\includegraphics[width=7cm,height=7cm]{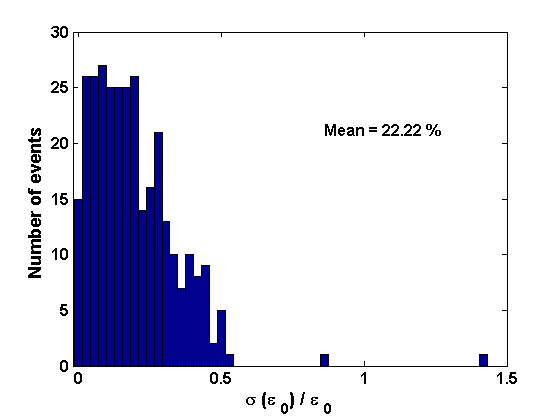}

\caption{Distribution of the estimated error of the electric field at the shower
footprint obtained from Monte Carlo calculations. A average relative
error $\sigma(\varepsilon_{0})/\varepsilon_{0}=22.22$ \% is obtained,
which will be used for the further analysis.}
\end{figure}

One of the other parameters for the radio-estimator of the energy
is the slope parameter $d_{0}$ whose distribution is presented in
Fig. 5. The peak value of the $d_{0}$ distribution is $156\: m$.
About $20$ \% of the events exhibit a $d_{0}$ greater than $400\: m$.
Two potential causes of this tail have been investigated: the malfunction
of some antennas and disruptions due to the configuration of the environment
near the antennas. After analysis, only $2$ \% of events were affected
by these disturbances \cite{key-34} and these were removed from
the data set.

\begin{figure}
\centering
\includegraphics[width=7cm,height=7cm]{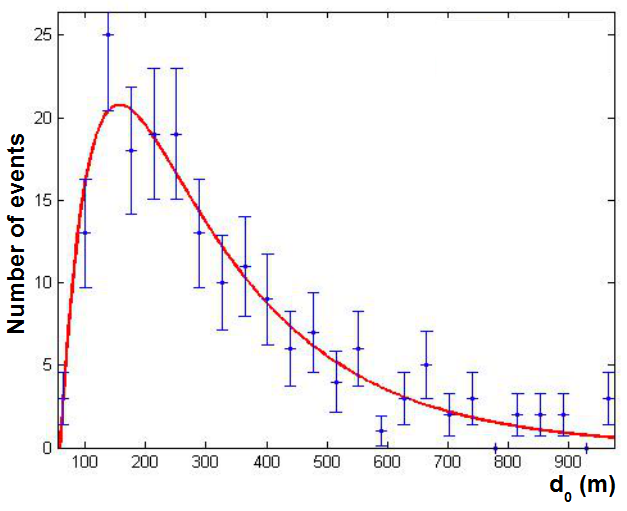}

\caption{Distribution of the slope parameter $d_{0}$ for the events ''radio''. The red line is to guide the eye.}
\end{figure}

As will be shown, the other valuable features of ``radio'' events
lie in the distribution of their arrival directions with respect to
the geomagnetic axis. It was shown in a previous work \cite{key-21}
that the electric field magnitude depends on the Lorentz force via
the vector cross product $\mathbf{v}\wedge\mathbf{B}$, where $\mathbf{v}$
is the vector velocity of the primary particle and $\mathbf{B}$ is
the earth's magnetic field (north-south oriented with a zenith angle
of $27\,{^{\circ}}$ at Nan\c cay). With our convention of angles,
the projection of the unitary vector of $\left|(\mathbf{v}\wedge\mathbf{B})_{EW}\right|$
on the east-west axis can be written as:

\begin{equation}
\left|(\mathbf{v}\wedge\mathbf{B})_{EW}\right|=\left|-\sin\theta\cos\phi\cos27-\cos\theta\sin27\right|
\end{equation}

This projection quantifies the influence of this process of emission.
The associated distribution of this quantity is depicted in Fig. 6.

\begin{figure}
\centering
\includegraphics[width=7cm,height=7cm]{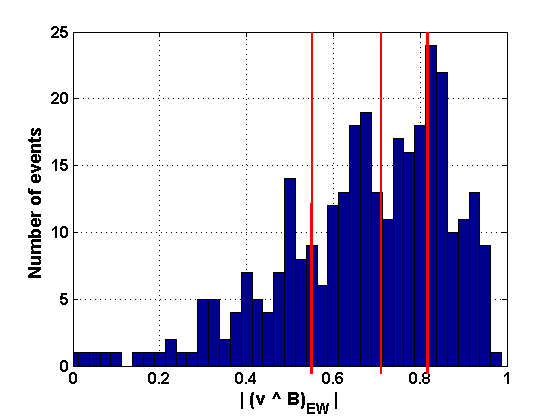}\caption{Distribution on the east-west axis of the quantity $\left|(\mathbf{v}\wedge\mathbf{B})_{EW}\right|$
for our sample of events. The mean of the distribution is located
at $0.67$. The windows used in the present study are marked in red.
Their corresponding means are $0.40$, $0.64$, $0.77$, and $0.88$.}
\end{figure}

\section{Electric field and energy correlation}

\subsection{Radio-estimator of the shower energy}

Basically, the electric field radiated should be linked to the energy
of the primary particle through the number of secondary charges produced
within the shower. According to our data, the only available observables
that can be directly related to the shower energy are based on the
RLDF parameters $(\varepsilon_{0},\, d_{0})$. The simplest assumption
suggests using the observables $\varepsilon_{0}$, the electric field
at the radio shower center, or $\intop\varepsilon_{0}\,\exp(-d/d_{0})\, dr=\varepsilon_{0}d_{0}^{2}$
, the integral of the RLDF. Because of difficulties in interpreting
huge values of $\varepsilon_{0}d_{0}^{2}$ for flat RLDFs, this latter
observable was rejected. The simplest quantity $\varepsilon_{0}$
was chosen to study the correlation with the shower energy estimator
$E_{P}$ given by the particle detector array.

Figure 7 shows the scatter plot of $\varepsilon_{0}$ versus $E_{P}$.
It exhibits a clear dependence of $\varepsilon_{0}$ on the primary
energy $E_{P}$ (to improve the clarity of the figure, the error bars
of $30\,$\% for $E_{P}$ and $22\,\%$ for $\varepsilon_{0}$ are
not represented). Indeed, the adjustment of this distribution with
the power law $\varepsilon_{0}\propto E_{P}^{\gamma}$ indicates an
almost linear relationship between both quantities (power coefficient
$\gamma=1.03$) in our detection range. Accordingly, we chosen to
fit the scatter plot distribution with the simple linear relation:

\begin{equation}
\varepsilon_{0}=\alpha\, E_{P}+\beta
\end{equation}

for which the parameters $\alpha$ and $\beta$ are determined by
minimizing, for all events $(\varepsilon_{0},\, E_{P})$, the quantity:

\begin{equation}
\chi^{2}=\underset{i}{\sum}\frac{[\varepsilon_{0}-(\alpha\, E_{P}+\beta)]^{2}}{[\sigma^{2}(\varepsilon_{0})+\alpha^{2}\sigma^{2}(E_{P})]}
\end{equation}

where $\sigma(\varepsilon_{0})$ and $\sigma(E_{P})$ are the experimental
standard deviations of the radio sigma $\varepsilon_{0}$ and the
energy $E_{P}$ respectively, deduced from CIC analysis and reported
on the measurement errors associated with each event.

\begin{figure}
\centering
\includegraphics[width=7cm,height=7cm]{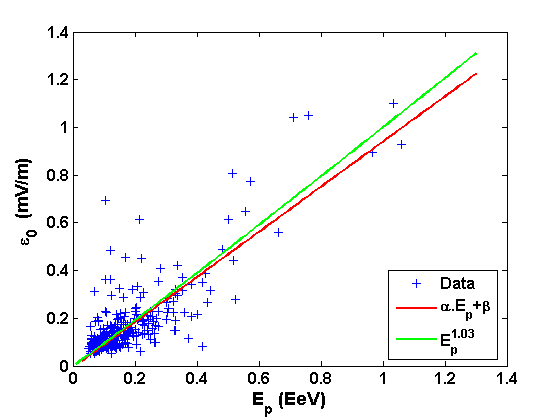}

\caption{Raw scatter plot of the electric field $\varepsilon_{0}$ at the shower
core versus the primary energy $E_{P}$ (for the readability of the
figure the error bars of $30$ \% for $E_{P}$ and $22$ \% for $\varepsilon_{0}$
are not represented). The adjustment of the data with the power law
$E_{P}^{1.03}$ is shown as a green line while the linear adjustment
is shown in red.}
\end{figure}

Inversion of the relationship enables the functional of the calibration
in energy $E_{0}$ of the radio signal to be extracted in the form:

\begin{equation}
E_{0}=\varepsilon_{0}/\alpha-\beta/\alpha
\end{equation}
\begin{equation}
=a\,\varepsilon_{0}+b
\end{equation}

Discrepancies between both observables $E_{0}$ and $E_{P}$ have
been investigated through the distribution of their relative differences,
defined as:

\begin{equation}
(E_{P}-E_{0})/E_{P}
\end{equation}

and its Standard Deviation (SD) given by (with $\mu$, the mean of
the distribution):

\begin{equation}
\sigma((E_{P}-E_{0})/E_{P})=\sqrt{\frac{1}{N-1}\sum([(E_{Pi}-E_{0i})/E_{Pi}]-\mu)^{2}}
\end{equation}

The typical form of distributions $(E_{P}-E_{0})/E_{P}$ is presented
in Fig. 8. The analysis of the width by a Gaussian fit was not considered
relevant because, as shown below, it is precisely the deviations from
the main peak of the events (the events in the tails) that reveal
the interesting effects.

\begin{figure}
\centering
\includegraphics[width=7cm,height=7cm]{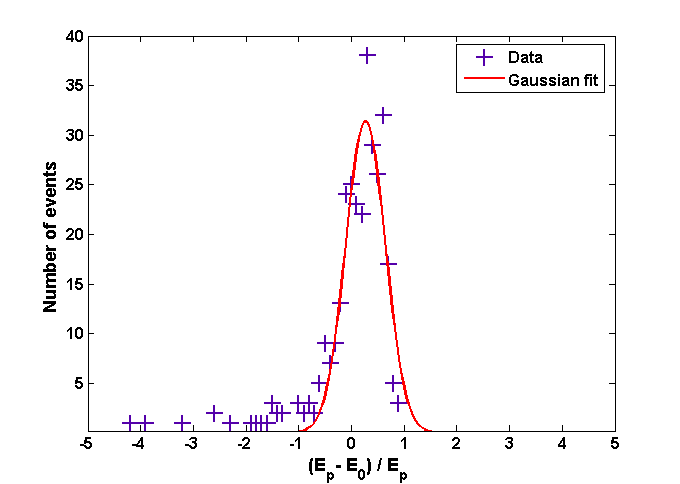}\caption{Distribution of the fractional difference between the energy estimated
with the scintillator array and the energy estimated with the antenna
array, obtained after calibration and for the full sample of events
(blue crosses). The standard deviation is $31$ \% with $c=0.95$
(see below). The red line shows a Gaussian fit made in the interval
$[-1,1]$.}
\end{figure}

\subsection{Correction factors}

In a previous article, we highlighted a significant asymmetry \cite{key-21}
of the EAS detection (in the east-west polarization) between the south
and north sectors of the sky. It was shown that the electric field
magnitude depends on the Lorentz force via the vector cross product
$\mathbf{v}\wedge\mathbf{B}$. To take this result into account, the
estimator $\varepsilon_{0}$ has been corrected by the inverse of
the projection of the Lorentz force along the east-west polarization
(Fig. 9). In other words, we assume that the observed electric field
$\varepsilon_{0}$ is proportional to the actual shower energy $E$
and that two detected showers of the same primary energy (and of the
same nature) will produce two different radio signals depending on
their angle with the geomagnetic axis, by following dependencies of
the form:

\begin{equation}
\varepsilon_{0}\sim E\left|(\mathbf{v}\wedge\mathbf{B})_{EW}\right|
\end{equation}
\begin{equation}
\varepsilon'_{0}\sim E\left|(\mathbf{v'}\wedge\mathbf{B})_{EW}\right|
\end{equation}

\begin{figure}
\centering
\includegraphics[width=7cm,height=7cm]{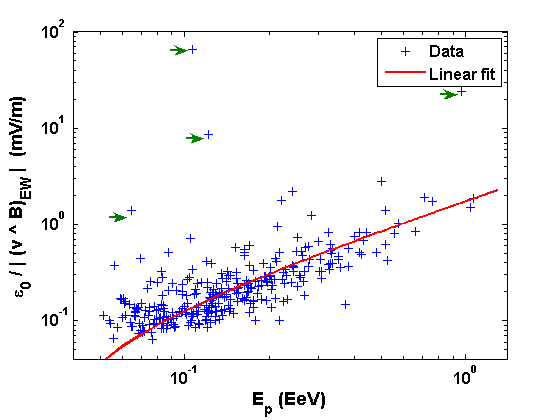}\caption{Scatter plot of the corrected electric field $\varepsilon_{0}/\left|(\mathbf{v}\wedge\mathbf{B})_{EW}\right|$versus
the primary energy $E_{P}$ . The linear adjustment is shown in red.
The green arrows show some of the typical over-corrected events. The
systematic analysis of their features indicates that all these events
have arrival directions near the earth's magnetic field direction
($\left|(\mathbf{v}\wedge\mathbf{B})_{EW}\right|<0.1$). (These points
are removed for the linear fit of the scatter plot).}
\end{figure}

The examination of the corresponding scatter plot $(E_{P},\, E_{0})$,
presented in Fig. 9, highlights a new phenomenon. With this correction,
the new locations of the events located close to the earth's magnetic
field direction ($\left|(\mathbf{v}\wedge\mathbf{B})_{EW}\right|<0.1$)
deviate significantly from the raw distribution of Fig. 7, exhibiting
unrealistic energy with the radio. In addition, with the assumption
of a purely geomagnetic emission process, these events should not
be radio-detected, particularly noting that their estimated CIC energy
shows nothing unusual. This suggests that the corrected amplitudes
are obviously overestimated, and that a further correction must be
introduced, to produce observable signals, especially when the Lorentz
contribution becomes weak. Moreover, in the case of a higher counting
statistics (larger counting time and detector surface), far above
the detection threshold, the rate of detected events around the direction
of the geomagnetic field should join to those of the other arrival
directions. The lack of a correction term additional would lead to
overestimate the energy and to strongly modify the energy spectrum.
To account from these remarks, we introduce a constant term $c$ in
our previous prescription, making the assumption that the observed
electric field $\varepsilon_{0}$ can also be proportional to the
actual shower energy $E$ by:

\begin{equation}
\varepsilon_{0}\sim E\, c
\end{equation}

the superposition of both contributions causing the dependence:

\begin{equation}
\varepsilon_{0}\sim E\left|(\mathbf{v}\wedge\mathbf{B})_{EW}\right|+E\, c
\end{equation}

leading to modify $\varepsilon_{0}$ toward:

\begin{equation}
\varepsilon_{0}\rightarrow\varepsilon_{0}/(\left|(\mathbf{v}\wedge\mathbf{B})_{EW}\right|+c)
\end{equation}

At first glance, the physical meaning of this additional contribution
is tricky, but it should be noted that, with the quantity $\left|(\mathbf{v}\wedge\mathbf{B})_{EW}\right|$ranging
from 0 to 1, $c=0$ corresponds to a purely geomagnetic contribution,
while a high $c$ value (e.g. $c>6$) should represent the case for
which the geomagnetic emission becomes negligible compared to the
second effect. Thus, a high $c$ value must induce a behavior similar
to that observed without correction.

The result is depicted in Figure 10, which presents the standard deviation
of the $(E_{P}-E_{0})/E_{P}$ distribution as a function of $c$ and
for several bins in $\left|(\mathbf{v}\wedge\mathbf{B})_{EW}\right|$.

\begin{figure}
\centering
\includegraphics[width=7cm,height=7cm]{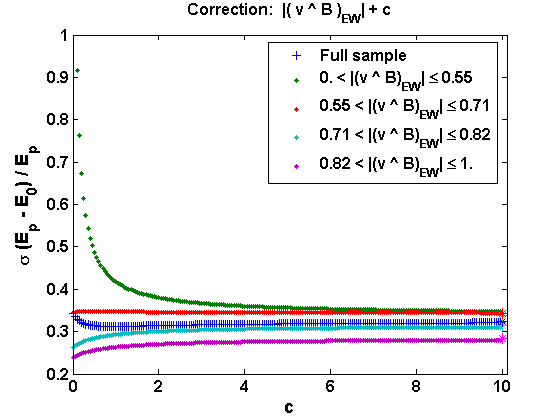}

\caption{Evolution of the Standard Deviation of the distribution $(E_{P}-E_{0})/E_{P}$
with the correction factor $1/(\left|(\mathbf{v}\wedge\mathbf{B})_{EW}\right|+c)$
as a function of $c$ and for several windows of$\left|(\mathbf{v}\wedge\mathbf{B})_{EW}\right|$.
The blue line represents the whole sample of events. The start located
$c=10$ show the SD obtained without applying a correction.}
\end{figure}

With the full sample of events (blue markers on figure 10), the result
seems monotonous with a weak minimum obtained around $c=0.95$. Concerning
the sub-samples, except for the two windows of lowest values in $\left|(\mathbf{v}\wedge\mathbf{B})_{EW}\right|$,
the raw values of the standard deviations (indicated by stars at the
right side of the figure 10) are bigger than when the corrections
are applied. The widths of the distributions depend clearly from the
direction of arrival of the showers with respect to the magnetic field
direction $\left|(\mathbf{v}\wedge\mathbf{B})_{EW}\right|$and with
$c$.

For the wide angles with the geomagnetic direction, the smallest dispersions
are achieved when $c=0$. This suggests that the geomagnetic correction
is significant and thus that this emission process is clearly dominant
in this region. On the contrary, for the events arriving near the
geomagnetic axis, large values of $c$ are called for. With regard
to these events, the low values of $c$ induce deviations that reflect
the larger gaps compared to the main peak of the distribution. (This
particular behavior explains why a Gaussian fit of the distributions
$(E_{P}-E_{0})/E_{P}$ was not used for the analysis.) For these events,
the contribution in $c$, directly correlated to the energy of the
shower, becomes significant, compared to the geomagnetic effect. The
remarkable values reached for the standard deviations are summarized
in Table 1 with their corresponding values of $c$.

\begin{table}
\begin{tabular}{|p{3.5cm}|p{2.5cm}|p{2.5cm}|p{2.5cm}|}
\hline $\sigma((E_{P}-E_{0})/E_{P})$&Correction $|(\mathbf{v}\wedge\mathbf{B})_{EW}|$ alone&Correction $|(\mathbf{v}\wedge\mathbf{B})_{EW}|$+0.95&No correction\tabularnewline
\hline Full sample (293 events) & 0.34 & 0.31 (minimum) & 0.32\tabularnewline
\hline $<|(\mathbf{v}\wedge\mathbf{B})_{EW}|>=0.4$ (70 events)  & 2.27  & 0.42  & 0.34\tabularnewline
\hline $<|(\mathbf{v}\wedge\mathbf{B})_{EW}|>=0.64$ (79 events)  & 0.34  & 0.35  & 0.34\tabularnewline
\hline $<|(\mathbf{v}\wedge\mathbf{B})_{EW}|>=0.77$ (68 events)  & 0.26  & 0.29  & 0.31\tabularnewline
\hline $<|(\mathbf{v}\wedge\mathbf{B})_{EW}|>=0.88$ (77 events)  & 0.23  & 0.26  & 0.28\tabularnewline
\hline
\end{tabular}
\caption{Standard deviation of $(E_{P}-E_{0})/E_{P}$ as a function of the correction factor $c$ varying arbitrarily from 0 to 10. In Gaussian approximation, the error on the standard deviation can be given by $\sqrt{2}\:\sigma^{2}/\sqrt{N-1}$ ($N\approx70$ for the subsets).} 
\end{table}

These results demonstrate that there is a strong correlation between
the radio signal, deduced using the peak amplitude of the extrapolated
electric field at the shower core, and the energy of the shower obtained
via the CIC method (although the latter is estimated without great
accuracy). The observed variations in function of $\left|(\mathbf{v}\wedge\mathbf{B})_{EW}\right|+c$
confirm the robustness of the correlation of the radio signal with
the actual energy of the shower and suggest strongly that the radio
signal is made of the mixture of several emission effects.

The association of the parameter $c$ to an existing electric field
was then investigated. Because of considerations of symmetry of the
electric field according to the axis of the radio signal, we have
focused on the contribution of an electric field component whose orientation
would follow the radio shower axis (eg. the one induced by the net
charge of the shower and which would remain basically proportional
to the energy of the shower). To address this possibility, we changed
the term $c$ by:

\begin{equation}
\varepsilon_{0}\sim E\left|(\mathbf{v}\wedge\mathbf{B})_{EW}\right|+E\, c\,\left|\sin\theta\,\sin\phi\right|
\end{equation}

The effect of this correction is depicted in Figure 11, which presents
the standard deviation of the $(E_{P}-E_{0})/E_{P}$ distribution
as a function of $c$ and for the same windows as in$\left|(\mathbf{v}\wedge\mathbf{B})_{EW}\right|$.
It clearly reflects a lower quality of the correlation and suggests
that the simple correction $\left|(\mathbf{v}\wedge\mathbf{B})_{EW}\right|+c$
captures better the experimental observations.

\begin{figure}
\centering
\includegraphics[width=7cm,height=7cm]{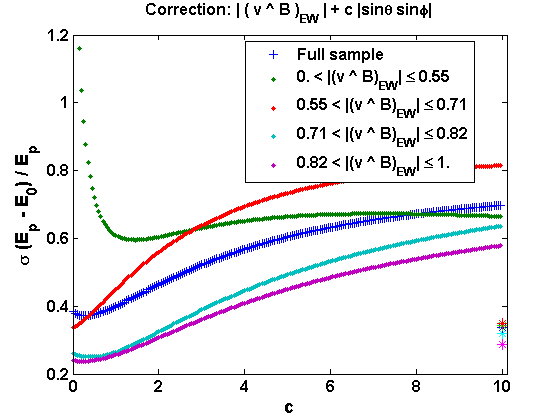}\caption{Evolution of the Standard Deviation of the distribution $(E_{0}-E_{P})/E_{P}$
with the correction factor $1./(\left|(\mathbf{v}\wedge\mathbf{B})_{EW}\right|+c\,\left|\sin\theta\,\sin\phi\right|)$
as a function of $c$ and for several values of $\left|(\mathbf{v}\wedge\mathbf{B})_{EW}\right|$.
The blue line represents the whole sample of events. The start located
at $c=10$ show the SD obtained without applying a correction.}
\end{figure}

A scalar emission mechanism was then searched for, particularly noting
that our observations were compatible with some recent theoretical
and experimental interpretations \cite{key-14,key-35,key-36}. For
example, the charge excess mechanism induces features that are in
qualitative agreement with our present observations: the induced emission
depends on the charges produced in the shower (and so from the energy);
its weight in the total emission process should increase when the
geomagnetic contribution, governed by $\mathbf{v}\wedge\mathbf{B}$,
decreases.

On another hand, by noting that $\left|(\mathbf{v}\wedge\mathbf{B})_{EW}\right|+c$
, can be rewriten $\left|(\mathbf{v}\wedge\mathbf{B})_{EW}\right|(1.+c/\left|(\mathbf{v}\wedge\mathbf{B})_{EW}\right|)$,
one could understand also the second term as an amplification effect,
inversely proportional to the Lorentz force. Taking into account the
forces acting on the secondary particles in the pancake, it could
suggest a coherence effect modulated by the Lorentz force. This process,
being similar to that caused by the deflection of a magnetic dipole
on particles in an accelerator beam line, could increase when the
secondary particles are less dispersed by the Lorentz force (i.e.
at small $\left|(\mathbf{v}\wedge\mathbf{B})_{EW}\right|$) during
their propagation.

\subsection{Calibration relationship}

Table 2 summarizes the calibration factors $(a,\, b)$ for our set-up,
with and without correction. Taking advantage of the full event feature
to deduce $\varepsilon_{0}$, the calibration can be expressed by:

\begin{equation}
E0=\frac{a}{(\left|(\mathbf{v}\wedge\mathbf{B})_{EW}\right|+c)}\varepsilon_{0}+b
\end{equation}

\begin{table}
\begin{tabular}{|c|c|c|}
\hline 
Full sample (293 events)  & a (EeV/mV) & b (EeV) \tabularnewline
\hline 
\hline 
No correction & $1.03\pm0.05$ & $0.0083\pm0.0004$\tabularnewline
\hline 
Correction$\left|(\mathbf{v}\wedge\mathbf{B})_{EW}\right|$ alone
(with $c=0$) & $0.59\pm0.06$ & $0.0230\pm0.0005$\tabularnewline
\hline 
Correction $\left|(\mathbf{v}\wedge\mathbf{B})_{EW}\right|+0.95$
(minimum) & $1.59\pm0.09$  & $0.0152\pm0.003$\tabularnewline
\hline 
\end{tabular}

\caption{Energy calibration coefficients in function of the the correction
factor in the cases ''without correction'', ''with the geomagnetic
effect alone'', ''with a scalar contribution (c=0.95) added to the
geomagnetic effect''.}
\end{table}

According to the different corrections, the slope parameter a of the
line varies by about $37$ \%. Similarly, for the value of the y-intercept,
a relative variation of b up to $36$ \% is observed. The latter could
be more related to the noise seen by the antennas which acts as an
offset in the amplitude of the radio-signal. As this noise contribution
is not constant, it can be only partial. Both parameters depend mainly
on the absolute calibration of the electric field antennas. At present,
it is not determined, in absolute terms, to better than a factor of
$2$. Nevertheless, the relative reference remains identical for each
antenna of the entire installation. Furthermore, the calibration depends
on the frequency bandwidth that is being analyzed, and on the frequency
response of the detection chain. This is, in principle, corrected
but the line parameters also depend on the frequency content of the
transient. Thus, it is likely that it will be necessary to readjust
the terms of the calibration relationship for each detector system.
This suggests that the implementation of calibration methods, accurate
enough and independent of the specific characteristics of the detectors,
will only be possible when a full model of the radio emission process
becomes available. The two energy spectra deduced from the CIC analysis
and calibration of radio, are presented in Figure 12. They indicate
a very satisfactory consistency between both observables of the energy
of the primary.

\begin{figure}
\centering
\includegraphics[width=7cm,height=7cm]{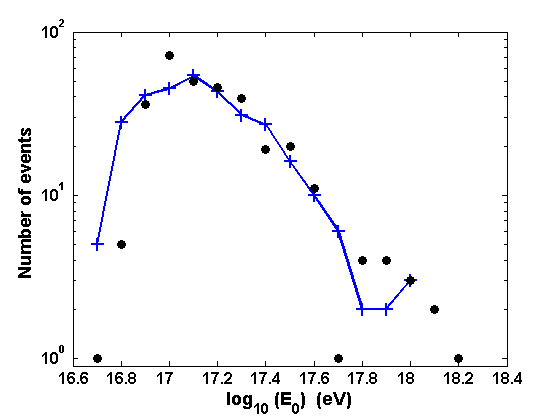}\caption{Comparison of energy spectra for events ``radio''. In blue, the
spectrum obtained by assigning to the events ``radio'' the energy
given by the particle detectors (cf. Fig 1, left panel); in black,
the spectrum obtained after calibration of the radio signal using
the correction $(\left|(\mathbf{v}\wedge\mathbf{B})_{EW}\right|+0.95)$.}
\end{figure}

\subsection{Energy resolution}

One of the most challenging aspects of the calibration procedure is
to assign an error to the observable $E_{0}$ provided by the previous
outputs. This has been addressed by studying the effect of $\sigma(E_{0})/E_{0}$
in a simulated distribution $(E_{P}-E_{0})/E_{P}$ ($\sigma(E_{P})/E_{P}$
being fixed), while noting that this procedure does not allow the
error on $E_{0}$ to be explored, on an event by event basis. A random
distribution in energy $E$ of the events is adjusted to the experimental
data (cf. left panel of Fig. 1). The pairs $(E_{P},\, E_{0})$ are
then derived. For each energy coming $E$, the associated energy $E_{P}$
assigned to the particle detector is taken randomly from a Gaussian
probability density function centered on $E$ with $\pm3\sigma$ and
with $\sigma(E_{P})/E_{P}$ fixed; similarly for the associated energy
$E_{0}$ of the radio-estimator with $\pm3\sigma$ and with the $\sigma(E_{0})/E_{0}$
chosen for the study. The distribution $(E_{P}-E_{0})/E_{P}$ is built
step by step and finally its standard deviation is calculated. No
correction with the geomagnetic effect is introduced because the Monte
Carlo provides $E_{0}$ directly. The results of the SD are combined
in Fig. 13 versus $\sigma(E_{0})/E_{0}$. With $\sigma(E_{P})/E_{P}=30\:\%$,
a relative error below $20\:\%$ is obtained for $E_{0}$.

\begin{figure}
\centering
\includegraphics[width=7cm,height=7cm]{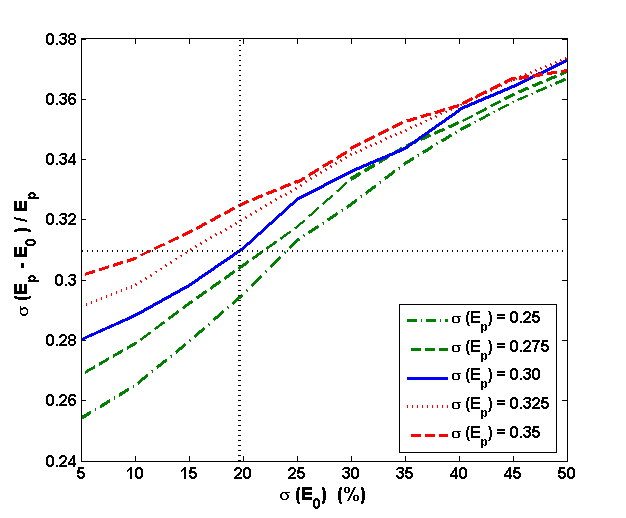}\caption{Reconstructed RMS of the simulated distributions $(E_{0}-E_{P})/E_{P}$
as a function of $\sigma(E_{0})/E_{0}$ and for $5$ values of $\sigma(E_{P})/E_{P}$
$(0.25,\;0.275,\;0.3,\;0.325,\;0.35)$. The solid blue line in blue
corresponds to the current energy resolution of $30\:\%$ of the particle
detector array.}
\end{figure}

One of the recurring issues with this kind of calibration is whether
the experimental result is the consequence of a strong hidden correlation
between the two energy observables, like that which might arise from
the uncertainty about the depth of the first interaction. Here, one
fundamental aspect could differentiate the radio from the particles
measurements. While the particle density is sampled at a fixed depth
of the shower, the radio signal integrates its development in the
atmosphere. In the energy range specifically studied here, the whole
development (including at the $X_{max}$ depth) is always seen by
the antennas. One can suppose, therefore, that this phenomenon will
not produce this bias in our ``radio'' observation.

\section{Conclusion}

We presented an attempt to estimate the energy of primary cosmic rays
using the measurement of the radio-electric field associated with
the air showers. We took advantage of the electric field $\varepsilon_{0}$
deduced at the core position of the shower through an exponential
Radio Lateral Distribution Function of the electric field. We showed
that a linear relationship can be found between this observable and
the inferred energy of the particle detectors, allowing the calibration
of the radio-detection method in the MHz range.

Taking into account previous results on the geomagnetic effect, we
then emphasized the utility to correct the electric field amplitude
by this effect. The study of the distributions of relative energies,
built from the radio and particles method, led us to search for adequate
relationships for the correction. The observed variations proved that
the correlation was robust and sensitive enough to be improved by
adjusting some parameters. This led us to consider the idea of a mix
of several emission effects. In the present state of the statistic,
a second correction proportional to the energy of shower (scalar contribution)
seems justified, as for instance coming from a coherence effect, although
our preliminary interpretations remain compatible with the recent
inclusion of new effects in the emission processes, like the charge
excess.

Combined with other detection performances (cost, ease of use, duty
cycle, etc.) our first estimate of the energy resolution (rather below
$20\:\%$) suggests that the method could be particularly attractive.
With the limited data set available, the absolute calibration of the
energy remains difficult even though it is largely compatible with
the measurement of the energy supplied by the particle detectors,
and therefore usable. In addition, it is not clear that we should
work in the referential of the particles or that provided by radio
signals. However, the principle of the correction doesn't depend of
the form of the Radio Lateral Distribution Function of the electric
field. We expect that the use of a more relevant RLDF still increases
the correlation, and thus that our present energy resolution correspond
to a pessimistic estimate for the radio-detection method.

More data should improve obviously the proposed approach, but there
is already an indication of a very interesting energy resolution and
a hope that the radio signal could directly depend of the simple parameters
considered. This could also be a significant advantage of the radio
method.\\

\end{document}